\documentclass[preprint]{aastex}
\usepackage{psfig}
\input{epsf}
\shortauthors{LYUBARSKY, EICHLER, AND THOMPSON}
\shorttitle{DIAGNOSING MAGNETARS WITH TRANSIENT COOLING}
\def\lambdabar{\protect\@lambdabar}
\def\lambdabar{%
 \relax \bgroup
\def\@tempa{\hbox{\raise.73\ht0
\hbox to0pt{\kern.25\wd0\vrule width.5\wd0
height.1pt depth.1pt\hss}\box0}}%
 \mathchoice{\setbox0\hbox{$\displaystyle\lambda$}\@tempa}%
{\setbox0\hbox{$\textstyle\lambda$}\@tempa}%
{\setbox0\hbox{$\scriptstyle\lambda$}\@tempa}%
{\setbox0\hbox{$\scriptscriptstyle\lambda$}\@tempa}%
\egroup}
\newcommand{\nn}{\mbox{} \nonumber \\ \mbox{} }

\begin{document}
\title{Diagnosing Magnetars with Transient Cooling}
\author{Yuri Lyubarsky, David Eichler}
\affil{Dept. of Physics, Ben-Gurion
University, Beersheba 84105, Israel; lyub,eichler@bgumail.bgu.ac.il}
\author{Christopher Thompson}
\affil{CITA, 60 St. George St., Toronto, ON, M5S 2W9, Canada;
thompson@cita.utoronto.ca}

\begin{abstract}
Transient X-ray emission, with an approximate $t^{-0.7}$ decay,
was observed from SGR 1900+14 over 40 days  following the
 the giant flare of 27 Aug 1998.  We calculate in detail the diffusion
of heat to the surface of a neutron star through an intense
$10^{14}-10^{15}$ G magnetic field, following the release of
magnetic energy in its outer layers.
We show that the power law index, the fraction of burst energy in
the afterglow, and the return to persistent emission
can all be understood if the star is composed of normal baryonic
material.
\end{abstract}

\section{Introduction}
It is now believed that soft gamma repeaters (SGR's) are
``magnetars"  - compact objects that have magnetic fields of order
$10^{15}$ Gauss (Duncan \& Thompson, 1992; Paczy\'nski 1992;
Thompson \& Duncan 1995, 1996). There has, however, been some
question raised as to the nature of the star that bears this magnetic
field, and the precise mechanism by which energy is released
during the bright X-ray flare of an SGR.  Alcock,
Farhi, \& Olinto  (1986), Cheng \& Dai (1998), Dar (1999),
Zhang, Xu, \& Qiao (2000) and Usov (2001) have  suggested
that SGR's are strange quark stars, motivated in part by the
super-Eddington luminosities of their giant flares.

There is evidence that the giant SGR flares involve the cooling of
a confined $e^\pm-$photon
plasma in an ultrastrong magnetic field.   For example, the light
curve in the 27 August 1998 giant flare terminated sharply some
400 s after the onset of the flare and can be fit accurately by
the contracting surface of such a cooling `trapped' fireball
(Feroci et al. 2001).   An account of the first stages of the
burst has been proposed in the strange quark star model (Usov 2001),
but not of this final drop in flux.

     How can one tell what a putative neutron star is really made of?
Eichler \& Cheng (1989, hereafter EC) suggested that afterglow from transient
energy release might be a way to thermally ``sound out" the nature
of the crust.  Heat released below the surface -- deep enough that
electron conduction dominates and neutrino losses can be neglected
-- would be mostly sucked into the star and radiated away as long
term, essentially steady emission.  Heating the crust at shallower
depths creates a thermal echo lasting about $10^4$ seconds or
less. However, it was also noted in EC that transient afterglow
could also occur on a timescale of months if it is  only a small
fraction of the energy, and could probably be observed only if the
energy of the outburst exceeded $10^{43}$ erg.  In this case, most
of the heat is sucked into the body of the star, but the surface
stays hot long enough to provide a transient tail of X-ray
emission.

   The inference of such deep heating during SGR bursts would provide
a diagnostic of how the bursts are triggered:  e.g., an
indication that a burst involves not only a rearrangement of the
magnetic field outside the star, but also a motion and deformation
of the crust itself.  Heat conducted into the surface from an
external fireball will produce afterglow immediately following
an SGR burst, and heat deposited near the base
of the crust will become visible on a timescale of $\sim 1$ year
(Thompson \& Duncan 1995, 1996).  On intermediate timescales one
is led to consider the release of magnetic energy at shallow
depths in the crust (at densities below neutron drip), which creates
an inverted temperature profile (temperature increasing outward).

      There are now observations of neutron star afterglow.
Rothschild,  Kulkarni,  \&  Lingenfelter (1994) discovered
persistent X-ray emission ($L_X \sim 7 \times 10^{35}$ erg/s) from
SGR 0526-66, and persistent emission at similar levels has been
discovered from the other 3 SGRs (Murakami et al. 1994; Hurley et
al. 1999; Woods et al. 1999). This luminosity is similar to that
obtained by averaging the release of $\sim 10^{45}$ ergs of
magnetic energy over an interval of 50 years between giant flares.
More recently, Woods et al. (2001) have reported a transient
brightening of SGR  1900+14 following the Aug. 27 giant flare,
which stays above the persistent emission for about 40 days.
During this time the luminosity in the 2-10 KeV X-ray band decays,
to a good approximation, as $t^{-0.7}$. The total emission in this band is
$\sim 10^{42}(D/10 {\rm kpc})^2$, about $10^{-2}$ of the observed flare energy.
The spectrum is non-thermal.

     In this letter, we calculate in detail the thermal echo emerging
from the crust of a magnetar, using realistic (magnetic) specific
heats and thermal conductivities.  This radiation may be
resonantly scattered in the magnetosphere  (Thompson, Lyutikov, \&
Kulkarni 2002) thereby obtaining a non-thermal spectrum.  The time
dependence, we suggest, is nonetheless established by the thermal
properties of the outer crust.  A companion paper will address the
cooling of a surface layer which is heated sufficiently to become
pair-loaded and non-degenerate, and compares the resulting light
curve with the $\sim 10^3$-s tail of X-ray emission detected
following the shorter Aug. 29 burst from SGR 1900+14 (Ibrahim et
al. 2001).

\section{Basic Assumptions}
 {\it Deposition of Heat.} We focus on the outer 500 m or so of the
 crust, within which the pressure of a (vertical) $\sim 10^{15}$ G
 magnetic field is comparable to or larger than the matter pressure.
 We assume that
 the crust is heated suddenly, within $10^4$ s.
 While  the results are not too sensitive to the assumed
 profile of heat deposition,  we can nevertheless envision a plausible
 mechanism:
In the outer crust,
 which has little  rigid strength, the toroidal field
relaxes to nearly a constant over any cylindrical segment
 of magnetic surface. The
 toroidal field changes during an outburst, in
 which there is shift within the deep crust (e.g. Thompson et
 al. 2002).  However, such relaxation probably occurs at
 different times and to different extents at
 different horizontal locations.  The
 resulting shear causes reconnection between neighboring magnetic
 surfaces,releasing much of the toroidal field energy locally.
 This dissipation is likely to be local. The heat density deposited
 on any magnetic
 surface is thus independent of depth. The results are shown below
 to be in fact somewhat flexible to this assumption.
%
 Large-scale shearing of the rigid crust by bulk magnetic stresses
is likely accompanied by the formation of smaller scale
dislocations
 and elastic deformations.  The enormous range of SGR burst
 energies (Gogus et al. 2000) gives indirect evidence that these
 deformations extend over a large range of scales.  The inference
 of bulk heating within the outer crust during an SGR burst therefore
 suggests that the crust does not merely fracture along large-scale
 faults, but is subject to a more continuous shear deformation.

We assume a deposition of  thermal energy density of $\sim 1\times
10^{25}$ erg cm$^{-3}$. This is near the maximum for which
neutrino losses can  be neglected, and it is comparable to the
ratio of the flare energy ($\ga 1\times 10^{44}$ ergs) to the
volume of the neutron star. Within the crust,  this energy density
is less than a percent of $B^2/8\pi$, but  {\it greater} than the
pre-existing thermal energy density at depths less than $z_{heat}
\sim 300$ m (for a likely internal temperature of $\sim 5-7\times
10^8 K$; Thompson and Duncan 1996). If deposited over the entire
surface and to a depth of  $\sim 500$ m, this energy density
implies a total energy of a few times the measured Aug 27
afterglow energy.

{\it Parameters of the Upper Crust.}
The super-strong magnetic field significantly affects the
structure of the upper crust. The Landau energy is relativistic in
a $\sim 10^{15}$ G magnetic field.  The $n^{th}$ level has an
energy
$E_n(b)=m_ec^2\sqrt{1+2bn}$,
where
$b\equiv {\hbar eB}/{m_e^2c^3}= B/({\rm 4.4\cdot 10^{13}\rm G})$
is the field strength in QED units.
The density distribution in the crust is found from the
equation of hydrostatic equilibrium,
$dP/dz=\rho g$.
In this  letter, we normalize the surface gravity to  $g=10^{14}$ cm/s$^2$
and neglect GR effects.  Below a depth of a few meters, the electrons
are degenerate and their density is
\begin{equation}\label{neval}
n_e=n_0\sum_{n=0}^{n_{max}}g_n\sqrt{(E_F/m_ec^2)^2-1-2nb}.
\end{equation}
Here $n_0=b/(2\pi^2\lambdabar_C^3)$,
 $\lambdabar_C=\hbar/m_ec$ is the reduced Compton wavelength, $n_{max}$ the
maximum Landau number available at the Fermi energy $E_F$, and $g_n$
the statistical weight of the Landau levels ($g_0=1$, $g_{n>0}=2$).
Assuming constant mass and charge numbers $A$ and $Z$, there is
a simple relation between depth $z$ and $E_F$
\begin{equation}\label{zval}
z=\frac{Z}{Agm_p}\left(E_F-m_ec^2\right)= 49\frac ZA
\left(\frac{E_F}{m_ec^2}-1\right)\, \rm m
\end{equation}
which does not depend on $B$. The density grows slowly,
$\rho\propto z$, while all the electrons populate only the
background Landau level. The first Landau level is achieved at the
depth $z_1=24.5(2Z/A)(6.7\sqrt{B_{15}}-1)$ m; at larger depths the
density grows $\propto z^3$, as when $B=0$.


Below a depth of a few meters, the heat is transferred by
degenerate electrons. We calculated the electron thermal
conductivity making use of the code developed by Potekhin (1999).
The electron thermal conductivity, $\kappa$, has a prominent peak
when $E_F$ is about the Landau energy. At larger density, $\kappa$
decreases, reaches a minimum when electrons become effectively
3-dimensional (at $z\sim 2z_1$) and then grows slowly, as in the
nonmagnetized case.  At small densities (at $z<z_1$), $\kappa$
rapidly decreases so that close to the surface the heat transfer
is dominated by radiation. Close to the surface, $\kappa$ is so
small that the heat resistance of the crust is dominated by the
upper few meters.

The specific heat of the magnetized electrons
experiences strong oscillations with depth; we calculate it
numerically, directly from the thermodynamic potential.
The ions form a liquid through most of the heated layer (below a density
$\sim 10^{10}$ g cm$^{-3}$);  we approximate their specific heat as
$C_{V,i}=3k_Bn_i$.

{\it Neutrino Cooling.}
Neutrino cooling in the outer crust of a neutron star is dominated
by pair annihilation $e^+ + e^- \rightarrow \nu + \bar\nu$
(e.g. Itoh et al. 1996).  Photo-emission and plasma
emission ($\gamma \rightarrow \nu + \bar\nu$) are subdominant.
We are most concerned with the region at $z \la 50$ m, where the
initial temperature may exceed $3\times 10^9$ K.  In this region
the electrons (and positrons) are largely confined to the lowest
Landau level, with thermal energy density\footnote{In this section,
we use units $k_B = c = \hbar = 1$.} $U_{\rm th} \simeq {1\over 12}eBT^2
(E_F/p_F)$.  Making use of the cross-sections of Loskutov and
Skobelev (1986), it is straightforward to write the cooling time
$t_\nu = U_{th}/\dot U(e^\pm\rightarrow\nu\bar\nu)$ as
\begin{eqnarray}\label{tnu}
t_\nu &=& {2^{3/2} \pi^{9/2}\over m_e^5 G_F^2\sum_i(C_{v,i}^2+C_{a,i}^2)}
\left({T\over m_e}\right)^{3/2}\,e^{(m_e+E_F)/T}\,f(E_F) \nn
&=& 4.1\times 10^4\,\left({T\over m_e}\right)^{3/2}\,e^{(m_e+E_F)/T}
f(E_F)\;\;\;\;\;{\rm s};
\end{eqnarray}
[$T\ll E_F, E_1(B)$].
The vector and axial-vector coupling constants sum to
$\sum_i(C_{v,i}^2+C_{a,i}^2) = 1.68$ over all three neutrino flavors.
The dimensionless function $f(p_F) = 3(m_e/E_F)^3$ for
relativistic electrons ($E_F \gg m_e$) and $f(p_F) = {1\over 16}m_e/E_F$
for non-relativistic electrons.

This cooling process is most important at shallow depths ($E_F < 1$ MeV),
but still carries away at most $\sim 10$ percent of the initial heat over
the first $10^4$ s.  For example,
$t_\nu = 2\times 10^5$ s for $T = 2.5\times 10^9$ K and $E_F = 1$ MeV.

\section{Solving the Heat Flow Equation}

We now solve the time-dependent heat flow equation
\begin{equation}\label{heatrans}
C_V\; \frac{\partial}{\partial t} T = \frac{\partial }{\partial
z} F;\qquad F= \kappa\;\frac{\partial}{\partial z} T.
\end{equation}
Here $z$ is measured downward.
[The steady heat transfer in the magnetars was considered recently
by Heyl \& Hernquist (1998) and Potekhin \& Yakovlev (2001).]
Although it is necessary to solve it numerically (see below), the
following analytical model, based on the above considerations is
illuminating.

We assume that the heat capacity is dominated by 3-dimensional,
relativistic electrons,
$C_V = C_{V,e} = \pi^2 n_ek_B^2T/E_F \equiv  K z^2 T$,
where $n_e = E_F^3/3\pi^2(\hbar c)^3$ is the electron density and
the constant $K = (Agm_pk_B)^2/3Z^2(\hbar c)^3$.  In the Coulomb liquid,
the thermal conductivity can be written as
$\kappa = (\pi/3)cE_Fk_B^2T/Ze^4\Lambda_c \equiv K^\prime zT$,
where $\Lambda_c \simeq 1.6$ is the Coulomb logarithm.
The initial temperature profile is related to the initial
thermal energy density $U_{th}$ through
$T(z,0)= (2U_{th}/Kz^2)^{1/2}\propto 1/z$,
where $U_{th} \propto B^2/8\pi$ is independent of $z$.  Thereafter,
a broad temperature maximum forms and propagates inward.
 Assuming the skin layer to be to zeroth approximation a perfect insulator,
the equation is linear in $T^2$ and it can be shown, via a Bessel
transform (EC),  that the solution at vanishing depth
$\epsilon$ below it is
\begin{equation}
T_{-} \equiv T(\epsilon,t) =
\left[\Gamma\left({1\over 3}\right)
\,\left(U_{th}\over 2K\right)\right]^{1/2}
\,\left(K^\prime t\over K\right)^{-1/3}.
\end{equation}
If the heat conductivity $\kappa$ had the same linear dependence
on $T$ in the skin layer, then a constant fraction of the heat
would escape through the surface, and the surface flux would scale
as $t^{-2/3}$.

The time-dependence of the observed 2-10 keV flux must be corrected for
the reprocessing of thermal surface photons (with luminosity $L_{th}$)
into a non-thermal spectral tail.  Assume the observed photon spectrum is
$dN/dE = AE^{-\Gamma}$ from energy $E_{th}$ to infinity.
The seed photon energy scales as $E_{th}\propto L_{th}^{1/4}$, and the total
flux of photons as $AE_{th}^{1-\Gamma}/(\Gamma-1) \propto L_{th}^{3/4}$.
Thus the normalization constant $A$ scales as
$E_{th}^{-1+\Gamma}L^{3/4}\propto L^{1+(\Gamma-2)/4}$. Because the
spectral index $\Gamma$  after the Aug 27 event was close to -2, the photon flux
in any given energy band is proportional to L.

{\it Numerical Solutions:}  We follow the cooling of an iron layer of
depth 0.5 km. In the outermost layers of the
crust,  radiation dominates  heat transfer. The outgoing thermal
flux is established within a ``sensitivity strip" (depth of
5-10 m) where the radiative and electron heat  conductivities are
comparable, and  where the overall heat resistance is a maximum
(Gudmundsson, Pethick, \& Epstein 1983; Ventura \& Potekhin 2001).
Within this strip, the opacity is predominantly free-free
absorption, and we used the fit of Potekhin \& Yakovlev
(2001) for the absorption coefficient.

In the outermost layer, the characteristic heat diffusion time,
$\tau \sim C_Vz^2/\kappa$, is so small that the heat flux is
nearly constant, $\partial F/\partial z=0$.  Thus we select an
outer `skin' zone where the steady state limit of (4)
was solved together with the equation of
hydrostatic equilibrium; this procedure gives the temperature at
the base of the skin zone, $T_{-}$, as an invertable function of
$F$. Essentially all the temperature drop occurs within the
sensitivity strip; below this layer, the electron thermal
conductivity grows rapidly with depth. We chose the bottom
boundary of the skin zone at $z\sim 30$ m where one can already
neglect thermal corrections to the equation of state.

In the skin zone, the equation of state was
chosen as a sum of the classical thermal pressure and the cold
degenerate electron pressure:
\begin{eqnarray}\label{pressure}
P=n_ek_BT&+&\frac 12
n_0m_ec^2\biggl[\frac{n_e}{n_0}\sqrt{1+(\frac{n_e}{n_0})^2}\nn
&+&
\log\left(\frac{n_e}{n_0}+\sqrt{1+(\frac{n_e}{n_0})^2}\right)\biggr].\nn
\end{eqnarray}
This expression has correct asymptotic behavior and provides 20\%
accuracy at $k_BT\sim E_F$. Beyond the skin zone, the full
nonsteady equations (4) 
were solved in a static density profile (1, 2)
subject to the upper boundary condition that $\kappa
\partial T/\partial z =F(T_{-})$, where $T_{-}$ is that found in
the previous time step.

\begin{figure}
\centerline{\psfig{file=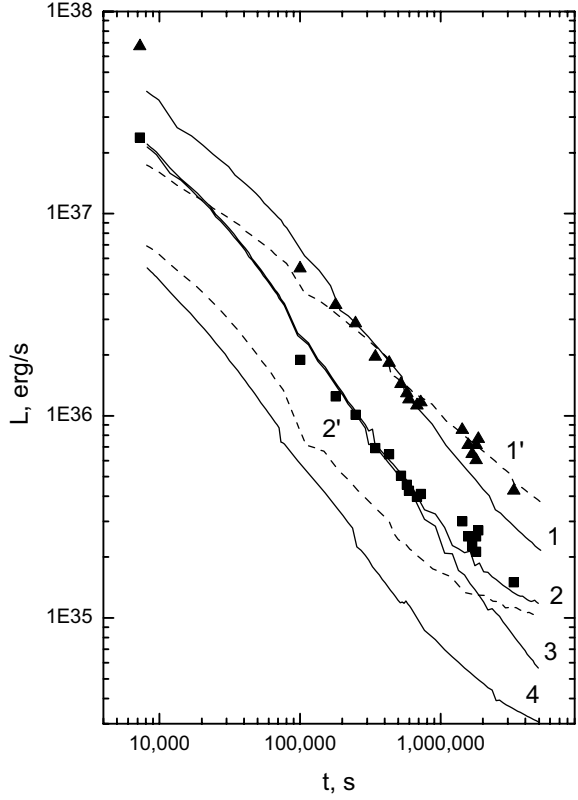,width=11cm,clip=} }
 \caption{Flux times $4\pi\times 10^{12}$ cm$^2$ as a function of time.
Curve 1 corresponds to $T_{max}=5\times 10^9$ K, $T_{int}=7\times
10^8$ K, $B=10^{15}$ G; curve 2 to $T_{max}=5\times 10^9$ K,
$T_{int}=7\times 10^8$ K, $B=3\times 10^{14}$ G; curve 3 to
$T_{max}=5\times 10^9$ K, $T_{int}=4\times 10^8$ K, $B=3\times
10^{14}$ G; and curve 4 to $T_{max}=3\times 10^9$ K,
$T_{int}=4\times 10^8$ K, $B=3\times 10^{14}$ G. The dotted curve
1' is for $B=10^{15}$ G and an initial temperature distribution $T=
5\times 10^9$ K at $z<30$ m, and $5 \times 10^9K (z/30{\rm m})^{-0.6}$ at
$z>30$ m; curve 2' is for $B= 3\times 10^{14}$G and $T= 5 \times
10^9$ K at $z<100$ m, and proportional to $z^{-2}$ at greater depths
until merging with the internal temperature of $7 \times 10^8$ K.
Data points are from Woods et al. (2001). Squares are normalized
to a distance of 9 kpc for SGR 1900+14 and triangle to 16
kpc.\label{image}}
\end{figure}

The calculated outgoing flux is plotted, as a function of time, in
Fig.\ 1. The initial temperature distributions in curves 1 through
4 correspond to uniform heat density, with $T$ decreasing inward
until it matched onto the initial (internal) value $T_{int}$. The
heat density was normalized by the temperature $T_{max}$ at the
bottom boundary of the skin zone. The remaining two curves show
that the results are rather robust to varying the initial
conditions. A slight "knee" occurs when the temperature maximum
passes the minimum of the electron conductivity (at a few $\times
10^4$ s for $B = 10^{15}$ G). Beyond this break, the light curve
has a slope which is independent of $B$, because the thermal
conductivity at greater depths approaches the $B=0$ value. An
"ankle" can occur beyond $10^6$ s, when the temperature maximum
merges with the interior region of almost constant temperature.

\section{Conclusions}

We find that the transient X-ray light curve of SGR 1900+14 in the
40 days following the Aug. 27 event is consistent with the
hypothesis that the SGR is a magnetar made of otherwise normal
material.  While there may be some freedom  in choosing the heat
deposition profile, the 40 day timescale is consistent with the
basic physics of an outer crustal layer which is supported by
relativistic degenerate electrons against gravity, and the heat
capacity and conductivity increase considerably with depth. The
power law
 index of the decay, though certainly inconsistent with a constant
 initial temperature,  is found to be weakly sensitive to the
 exact initial temperature profile:  on timescales more than
 a few days, the deeper layers are in any case cooled by inward
 conduction.  Qualitatively, this causes all but
$\sim 20\%$ of the heat to be sucked into the star and reradiated only over
much longer timescales as surface X-ray emission or neutrinos.
The resulting transient afterglow emission is $\sim 1$ percent of the
flare energy, as observed (Woods et al. 2001), if the initial
thermal energy density in the crust is comparable to the ratio
of the flare energy to the volume of the neutron star.
This is also consistent with the observation that the time
integrated luminosity of the SGR is dominated by steady emission
rather than by the decaying post-burst flux.

While we have not disproved other compositions for the SGR -- e.g. a quark
star, which would have a much more homogeneous density -- the
question that arises is whether the thermal response of such
an object would be similar.  One expects the ratio of magnetic
and material pressures to be more nearly constant within the quark matter,
than in the stratified crust of a neutron star.   A power-law cooling
behavior can still be obtained on short timescales, but at the cost of
introducing a new scale to the problem:  the heat must be deposited only
to a finite depth in the quark matter.   Shallow heating of homogeneous matter
with a free escape boundary condition also implies that most of the
heat escapes the surface;  the depth of heating must therefore be
adjusted to give $\sim 1$ percent of the flare energy.
An insulating envelope at the surface could reduce
this problem, but its thickness would have to be adjusted to give a
conduction time less than $\sim 10^4$ s.  In that case, the temperature
at the outer boundary of the quark matter declines as
$T_- \sim t^{-1/3}$, and the surface X-ray flux as $T_-^n \sim t^{-n/3}$
with $n \sim 2-3$.
(Here we take into account that $\kappa$ is approximately independent
of temperature in the quark matter, and $C_V \propto T$; Heiselberg
\& Pethick 1993.)  Finally, a power-law behavior can also be obtained
from deep heating, but only on very long timescales comparable to the
cooling time of the star as a whole.


\acknowledgements We thank A.\ Potekhin and D.\ Yakovlev for
invaluable assistance, and Peter Woods for conversations.
We also thank the Institute for Theoretical
Physics, UCSB (NSF grant PHY 99-0749) for its support during the
workshop on ``Spin, Magnetism, and Rotation in Young Neutron
Stars,'' where this work was initiated.  We acknowledge with
gratitude  the Arnow Chair of theoretical Astrophysics, an
Adler Fellowship via the Israel Science Foundation, support from
the Israel Ministry of Absorption and  from a seed grant
from Ben Gurion University.  The work of CT is supported by the
NSERC of Canada.

\end{document}